%% file: 3675.YMExtraInv.tex
\font\caps=cmcsc10 at 12pt
\font\eightrm=cmr8 at 8pt
\documentclass{oxarticle}
\usepackage{amsmath, amssymb}
\input{Johnmacros15}
\newcommand{\articlenumber}{\LT3675.YMExtraInv.tex}

\proofmodefalse
\begin{document}
\begin{center}

\vspace*{1in}
{  \huge 
Extraordinary Invariants  \\are Seeds
that Grow \\ Interacting Theories
Out of Free Theories\\}

\renewcommand{\thefootnote}{\fnsymbol{footnote}}
%\footnotetext[1]{~here we have a footnote.}
\renewcommand{\thefootnote}{\arabic{footnote}}
\vspace{1cm}
{\caps John A. Dixon\footnote{jadixg@gmail.com}\\Toronto, Canada}
\\[1cm] 
%{\it No Fixed Address }
%\vfill

{\bf Abstract}
\end{center}

{\EI s are elements of the BRST Cohomology Space which
are  irrevocably dependent on Zinn sources.  The existence of an \EI\ means that  the symmetry is broken in that sector, and that the field equations can almost rescue the invariance.  Adding the \EI\ to the action results in a new theory with constraints on the starting theory. 

So \EI s are seeds from which a theory can grow.  For a simple example,  it is shown in this paper how
 \YM\ is implicitly contained in the BRST Cohomology of  Free Gauge Theory.
It comes from an \EI\ which can be added to the free gauge action.  The Jacobi Identities are generated by requiring that the \PB\ be zero.

Since the mechanism is a general one, it can be used to construct new theories.  Some of these, for example in Supersymmetric theories, have not yet been noticed  using other methods.

%\Large
\section{How can we construct interesting new Actions in  Quantum Field Theory?}

Quantum field theory has been very successful, notably for the Standard Model \ci{SMexpmt}.   Over a long period, spontaneously broken Yang-Mills gauge theory was gradually developed, starting from quantum electrodynamics, and both of these were incorporated into the Standard Model \ci{history}. 
Recently it appears that yet another confirmation of this success is arising from the discovery of the Higgs particle \ci{higgs}.  Supersymmetry is an important development too, but its application to the real world is still very problematic, and possibly non-existent \ci{susyconf}. String and Superstring theory are also very interesting, but again their applicability is far from understood in a satisfactory way \ci{stringconf}.

Of course, nobody believes that the Standard Model is the end of the story.  The problem is that most efforts to extend the \SM\ result in something that is experimentally wrong.   The development of new theories, like supersymmetry and superstring theory, has taken a long time.   So it is natural to wonder how one can come up with new theories to test against the experimental results. Is it conceivable that there is a simple way to generate new theories that have some good features? Surprisingly, the answer seems to be yes.

The present paper suggests a new, simple, easy  and  pedestrian approach to searching for new theories, and new variations on theories. One starts with a free action.  Then there is a procedure to generate a special set of interacting actions from it.  Moreover, the interacting actions will have a number of desirable features.

It is well known that our understanding of many theories has been improved by the methods of BRS and T
 \ci{BRSTcohom,Tyutin,ZJ}, and their use by many authors.  In particular the application of BRST cohomology  has  helped to clarify renormalization, physical states,  and anomalies in many contexts, both in quantum field theory \ci{incurablewithoutYM,henneaux}, and in string theory \ci{GSW}.   The  new approach in this paper to finding theories comes directly from the BRST formalism.  The method is to start with a free quantum theory and  look  for its {\bf `\EI s'}. Briefly,  an \EI\ is an Invariant in the BRST cohomology
which is `irrevocably dependent' on a Zinn source in its expression\footnote{This means that the $\d_{\rm Little} $ variation of the pure field part of the \EI\  is proportional to   
the equations of motion of the theory.  An example of this is set out in  equation (\ref{lorentzinvei}) of section \ref{extrainvsection} below.}.  

  The approach is easily illustrated for the well-known example of Yang-Mills theory.
If one looks at the BRST cohomology of free gauge theory, one can see that the interacting \YM\ is `lurking there'.  It takes the form of the
 `\EI' written below in equation (\ref{lorentzinvei}).  This is merely a curiosity for Yang-Mills theory,
but in some other theories, notably chiral scalar supersymmetric theories in 3+1 dimensions, the \EI s  seem to take on a more important role, which has yet to be fully worked out \ci{DASSM}.  Also, there are many free actions for which nothing is known about the \EI s. 

So the present paper is an introduction to the method in a simple and familiar context.

  Cohomology calculations have usually been done for interacting theories rather than free theories, so the 
\EI s of free theories have been ignored up to now.  Moreover,  \EI s  frequently have uncontracted Lorentz indices, so that they are  not Lorentz scalars, and so they have usually been ignored for that reason also, since research has been concentrated on analyzing actions rather than finding new ones. However the present results show that the analysis of \EI s is actually very important.

   In the case of \YM, we can find  an \EI\ in the free theory which is also a Lorentz scalar, so \YM\ is a particularly simple case and it constitutes a nice example of what happens. In more tricky cases, one can usually couple an \EI\ to an appropriate field to make a Lorentz scalar. This 
 provides a route to a  new action that also involves the new field. 

 In a rather weak way, the method recalls the old idea of the `Bootstrap' \ci{Bootstrapref}, because the theory `pulls itself by its own bootstraps' from a seemingly simple,  empty and  innocuous,  free theory to a significant, physically interesting and   interacting theory.  In the process, it uses the nice properties of the BRST formalism, which incorporates symmetry, physical states, elimination of redundant or unphysical states, the equations of motion, the possibility of anomalies, the elimination of unitarity ghosts and the elimination of gauge degrees of freedom and, possibly,  other things that have not yet been understood.
Of course, a `Bootstrap' needs to be non-linear.
This requirement is supplied by the very simple (but extremely non-trivial) quadratic non-linearity of the \PB.  All of this becomes clearer when one takes a simple  example, and so that is what we will do.

\section{Extraordinary Invariants in Gauge Theory}

\la{extrainvsection}

Every free quantum field theory, with any symmetry at all, can be formulated in such a way that it yields zero for some Grassmann Odd Poisson Bracket, with a corresponding `square root', which is the  nilpotent BRST differential $\dB$. Here we shall not try to be general however.  We shall just discuss pure free gauge theory in four dimensions, without any matter, except that we add an index a to the  gauge bosons  $A_{\m}^a$ so that we keep the possibility of having more than one of them.

For free gauge theory, the Poisson Bracket\footnote{The Zinn sources $\S_{\m}^a$ and $K^a$ have ghost charge -1 and -2 respectively, and the ghost $\w^a$ has ghost charge 1.  We try to use Greek letters for Grassmann odd quantities like 
$\S_{\m}^a,\w^a$ and ${\d}_{[\cal A]}$, and Latin letters for Grassmann even quantitites like $A_{\m}^a,K^a$.  } is: 
\be
{\cal P} \lt [ {\cal A}\rt ]=\int d^4 x 
\lt \{
\fr{\d {\cal A}}{\d \S^{a\m}}
\fr{\d {\cal A}}{\d A^{a}_{\m}}
+
\fr{\d {\cal A}}{\d K^{a}}
\fr{\d {\cal A}}{\d \w^{a}}
\rt \}
\la{PBA}
\ee
and the action is
\be
{\cal A}_{\rm Free} 
=\int d^4 x 
\lt \{
-\fr{1}{4}F^{a \m \n} 
F^{a}_{ \m \n} 
+
\S^{a \m} \pa_{\m} 
\w^{a}
\rt \}
\la{Freeaction}
\ee
where
\be
F^{a}_{\m \n} 
=\pa_{\m} A^a_{\n} 
-\pa_{\n} A^a_{\m} 
\ee
Note that the source $K^a$ does not appear in the action.  However, we keep it in the \PB\ for completeness\footnote{We could add another variation,  for example the exterior derivative $\x \pa$,  so that $K^a$  would appear in the action in the term  $K^a \x \pa\w^a $.}.
We also define the corresponding BRST  operator:
\be
{\d}_{[\cal A]}=\int d^4 x 
\lt \{
\fr{\d {\cal A}}{\d \S^{a\m}}
\fr{\d  }{\d A^{a}_{\m}}
+
\fr{\d {\cal A}}{\d \w^{a}}
\fr{\d  }{\d K^{a}}
+
\fr{\d {\cal A}}{\d A^{a}_{\m}}
\fr{\d  }{\d \S^{a\m}}
+
\fr{\d {\cal A}}{\d K^{a}}
\fr{\d  }{\d \w^{a}}
\rt \}
\ee
It is easy to see that if the \PB\ for a given ${\cal A}$ yields zero, then the corresponding 
${\d}_{[\cal A]}$ is nilpotent:
\be
{\cal P} \lt [ {\cal A}\rt ]=0
\Ra
({\d}_{[\cal A]})^2=0
\ee

It is easy to verify that both of these are true for the free action 
 (\ref{Freeaction}): 
\be
{\cal P} \lt [ {\cal A}_{\rm Free}\rt ]=0
\Ra
({\d}_{[{\cal A}_{\rm Free}]})^2=0
\ee

Let us use the simpler notation
\be
{\d}_{[{\cal A}_{\rm Free}]}
\equiv 
\d_{\rm Free}
\ee
Then we find that 
\be
\d_{\rm Free}
=
\d_{\rm Field\;Equation}
+
\d_{\rm Zinn}
+
\d_{\rm  Little}
\la{BigBRSTem}
\ee
where
\be
\d_{\rm Field\;Equation} = 
\int d^4 x
\lt \{
\pa^{\n}
\lt (
 \pa_{\n} A^a_{\m}
- \pa_{\m} A^a_{\n}
\rt ) \fr{\d}{\d \S^a_{\m}}
\rt \}
\la{fieldeqem}
\ee
and
\be
\d_{\rm Zinn} = 
\int d^4 x
\lt \{
 \pa^{\,\m} \S^a_{\m}
 \fr{\d}{\d K^a}
\rt \}
\la{zinnotherem}
\ee
and
\be
\d_{\rm Little}
 = 
\int d^4 x
\lt \{
 \pa_{\m} \w^a
 \fr{\d}{\d A^a_{\m}}
\rt \}
\la{LittleBRSTem}
\ee

 It is simple to calculate the BRST cohomology of such a free  theory for low dimensional integrated polynomials in the fields and sources. For the simple cases we are interested in here, that can be done by writing down all the possible \LIP s in the sector of interest.

Invariants are ghost charge zero  \LIP s that are in the cohomology space of the theory.   There are two kinds of Invariants that one may find:
\ben
\item
There may be `\OI s'. 
  These are Invariants that can be written purely in terms of the gauge fields.
Here is an example of an \OI.  This is a Lorentz tensor, so it is not Lorentz invariant.  One could make it Lorentz invariant by contracting the indices with 
$\d_{\rho}^{\lambda}$:
\be
{\cal O}^{\rho}_{\lambda} 
=
t^{abcd}
\fr{1}{m^4}\int d^4 x \lt \{
F^a_{ \m \n}
F^{b \m \n}
F^{c \n \rho}
F^d_{ \n \lambda}
\rt \}
\la{firstLIP1}
\ee

\item
There may be `\EI s'.  These are Invariants that depend crucially on the presence of Zinn sources in the Invariant.
\ben
\item
Here is an example of an \EI.  It is a  Lorentz vector:
\be
{\cal E}_{\m} 
=
\int d^4 x 
E_{\m} 
=
t^{ab}\int d^4 x \lt \{
\S^{b}_{ \m} \w^{a}
+
A^{ a \n} \lt (
 \pa_{\n} A_{\m}^{b}
- \pa_{\m} A_{\n}^{b}
\rt )
\rt \}
\la{firstLIP1}
\ee

It is easy to verify that 
\be
\d_{\rm Free} {\cal E}_{\m} =0
\ee
It is also easy to see that there is no integrated local polynomial 
${\cal B}_{\m}$ such that
\be
{\cal E}_{\m} =
\dF{\cal B}_{\m} 
\ee
This  means that ${\cal E}_{\m} $ is indeed in  the non-zero local cohomology space of $\dF$. Also,  it is easy to show that there is no possible 
local ${\cal Q}_{\m} $ such that the expression ${\cal E}_{\m}^{' }=
{\cal E}_{\m}  + \dF {\cal Q}_{\m} $ is free of the Zinn sources, and  so it follows that ${\cal E}_{\m}$ is an \EI.
\item
 Here is an example of  another \EI\footnote{ Note that (\ref{lorentzinvei}) is constructed from (\ref{firstLIP1}) by adding another vector field.   However (\ref{lorentzinvei}) also needs the extra term $\fr{-1}{2}
K^{a}
\w^b
\w^c$, which uses (\ref{zinnotherem}) to compensate  the variation, from (\ref{LittleBRSTem}), of the term in (\ref{lorentzinvei}) which contains $\S_{\m}^a$.}.  It is  a Lorentz scalar: 
\be
{\cal E}
=
\int d^4 x
\lt \{
\S^{a}_{ \m}A^{b \m}
 \w^{c}
-
\fr{1}{2}
\lt (
 \pa_{\m} A_{\n}^{a}
- \pa_{\n} A_{\m}^{a}
\rt )
A^{b \m}
A^{ c \n} 
-\fr{1}{2}
K^{a}
\w^b
\w^c
\rt \}
f^{abc}
\la{lorentzinvei}
\ee
Again, in this case, it is easy to verify that 
\be
\d_{\rm Free} {\cal E} =0
\ee
provided that $f^{abc}$ is a totally antisymmetric tensor. It is also easy to verify  that there is no integrated local polynomial 
${\cal B} $ such that
\be
{\cal E}  =
\dF {\cal B} 
\ee
which means that ${\cal E}  $ is indeed in the non-zero  local cohomology space of $\dF$. Also, it is easy to verify  that there is no possible 
local ${\cal Q}  $ such that the expression ${\cal E}^{' }=
{\cal E}   + \dF {\cal Q}  $ is free of the Zinn sources, and  so it follows that ${\cal E} $ is an \EI.
\een

\een

It is common practice to add  Ordinary Invariants, provided that they are Lorentz scalars, to the relevant free theory to get 
another (usually interacting) theory.  
But what happens if one tries to add an \ELI\ to the free action?
We shall see that adding an \ELI\ generates a related but different theory.  In some sense, the BRST Cohomology of the free theory here is `aware of and prepared for' the possibility of extending the theory to \YM, and that arises from (\ref{lorentzinvei}).

\section{Construction of \YM\ using the \ELI}

Let us verify that $\dF$ acting
 on (\ref{lorentzinvei}) yields zero.  This will 
also give us some insights.

  Integrating by parts yields the following constraints:
\be
\dB {\cal E} =0 \Ra
\eb
\fr{1}{2}\int d^4 x  
 f^{abc}
\lt ( \pa^{  \n} A^{a \m}-\pa^{  \m} A^{a \n}
\rt ) \w^b \lt (
 \pa_{\n} A_{\m}^{c}
- \pa_{\m} A_{\n}^{c}
\rt )
 =0
\la{symgaugeproblem}\ee

This equation is true if and only if  :
\be
f^{abc} = - f^{cba}
\la{symgauge}
\ee
A simple way
 to satisfy this constraint, is to ensure that $f^{abc}$ is a totally antisymmetric tensor\footnote{There are also trivial solutions of this with extra U(1) gauge fields.}.   
The next thing to notice is that if we add this expression ${\cal E} $ to the free action, and define:
\be
{\cal A}_{1}
={\cal A}_{\rm Free}
+ {\cal E} 
\la{eadded}
\ee
then the Poisson Bracket is no longer zero, but it does reduce to:
\be
{\cal P}\lt [{\cal A}_{1}\rt ]
=
{\cal P}\lt [{\cal E}\rt ]
\ee
This is a simple consequence of 
\be
{\cal P}\lt [{\cal A}_{\rm Free}\rt ]
=0
\ee combined with:
\be
\dF {\cal E}
=0
\ee
However in general 
\be
{\cal P}\lt [{\cal E}\rt ]
\neq 0
\la{pofenotzero}
\ee
To arrive at a new theory, starting from 
(\ref{eadded}), which yields zero for the BRST Poisson  Bracket, we must add something more to get 
\be
 {\cal A}_{\rm Complete}
=
{\cal A}_{\rm Free}
+ {\cal E} 
+
 {\cal A}_{\rm Completion}
\la{fullmonte}
\ee
such that the Poisson Bracket yields zero for the completed Action $ {\cal A}_{\rm Complete}$:
\be
 {\cal P} \lt [ {\cal A}_{\rm Completion}\rt ]
=0
\ee
Given the above, this means that the new term 
${\cal A}_{\rm Completion}$ must satisfy:
\be
{\cal P}\lt [{\cal E}\rt ]
+
\dF  {\cal A}_{\rm Completion}
=0
\la{compconstraint}\ee
and also:
\be
{\cal P}\lt [{\cal A}_{\rm Completion}\rt ]
=0
\ee
To proceed further we must evaluate the expression
${\cal P}\lt [{\cal E}\rt ]$ and see whether there is a solution for (\ref{compconstraint}). This is easy: 
\[
{\cal P}_{\cal E}=
\int d^4 x 
\lt \{
\fr{\d {\cal E}}{\d \S^{a\m}}
\fr{\d {\cal E}}{\d A^{a}_{\m}}
+
\fr{\d {\cal E}}{\d K^{a}}
\fr{\d {\cal E}}{\d \w^{a}}
\rt \}
\]\[=
\int d^4 x 
\lt \{
f^{abc}A^{b \m}\w^c
\lt (
f^{ade}
\lt [
-   
\pa_{\n}
(A^{d}_{ \m}
A^{ e \n} )
-\pa_{\n} A_{\m}^{d}
A^{ e \n} 
+
 \pa_{\m} A_{\n}^{d}
A^{e \n}
\rt ] +
f^{dae} \S^{d}_{\m}
\w^e
\rt )
\rt \}
\]
\be 
-
\int d^4 x 
\lt \{
\fr{-1}{2}
f^{abc}
\w^b
\w^c
f^{ade}
\lt (
-\S^{d}_{\m}
A^{e \m}
-
K^{d}
\w^e
\rt )
\rt \}
\la{poissonofextra}
\ee
There are three things that happen here:
\ben
\item
We  need to impose additional constraints on the tensor $f^{abc}$ so that this expression 
in (\ref{poissonofextra}) is in the image of $\dF$: 
\ben
\item
It is clear that the following can never arise from $\dF$:
\be
f^{abc}
\w^b
\w^c
f^{ade}
K^{d}
\w^e
\la{jacobihere}\ee
\item
It is also clear that the following can never arise from $\dF$:
\be
\int d^4 x 
\lt \{
f^{abc}A^{b \m}\w^c
f^{dae} \S^{d}_{\m}
\w^e
-
\fr{1}{2}
f^{abc}
\w^b
\w^c
f^{ade}
\S^{d}_{\m}
A^{e \m}
\rt \}
\la{jacobihereagain} 
\ee
\item
So to proceed we must set both of the above to zero.
Fortunately, this is a logical thing to do, since
(\ref{jacobihere}) and (\ref{jacobihereagain}) 
are both satisfied if the Jacobi Identity for the tensor $f^{abc}$  is true.
\een
\item
So these constraints mean that the
 term (\ref{poissonofextra}) reduces to the following term:

\[
\int d^4 x 
\lt \{
f^{abc}A^{b \m}\w^c
\rt \}
f^{ade}
\lt \{
-   
\pa_{\n}
(A^{d}_{ \m}
A^{ e \n} )
-\pa_{\n} A_{\m}^{d}
A^{ e \n} 
+
 \pa_{\m} A_{\n}^{d}
A^{e \n}
\rt \}
\]\be
=
\int d^4 x 
\lt \{
f^{abc}A^{b \m} \pa^{\n}  \w^c
f^{ade}A^{d}_{\m}
A^{e}_{ \n} 
\rt \}
\ee
where we use the circular symmetry from the Jacobi identity on the indices (bde) to write:
\be
\int d^4 x 
\lt \{
f^{abc}f^{ade}
A^{b \m}\w^c
 \pa_{\m} A_{\n}^{d}
A^{e \n}
\rt \}
\eb
=
-\int d^4 x 
\lt \{
\lt ( f^{adc}f^{aeb} + f^{aec}f^{abd} \rt )
A^{b \m}\w^c
 \pa_{\m} A_{\n}^{d}
A^{e \n}
\rt \}
\ee

Now we need to try to find a form which satisfies (\ref{compconstraint}).
Fortunately there is such a form:
\be 
{{\cal A}_{\rm Completion}}
=
\fr{-1}{4}\int d^4 x 
\lt \{
f^{abc}A^{b}_{\m} A^{c}_{\n} 
f^{ade} A^{d \m}
A^{e \n} 
\rt \}
\ee
and it is easy to verify that 
 (\ref{compconstraint})
is true.
\item
Finally we need to confirm that
\be
 {\cal P} \lt [ {{\cal A}_{\rm Completion}}\rt ]
=0
\ee
This is simple because ${{\cal A}_{\rm Completion}}$ does not have any Zinns in it, and ${\cal P} \lt [ {\cal A} \rt ]$ must be zero on any expression ${\cal A}$ without Zinns.

\een

\section{Conclusion}

\la{conclussection}

The final result for (\ref{fullmonte}) is the usual Yang Mills action\footnote{We have ignored the gauge fixing term and the ghosts--this can be done because of the usual methods used to derive the action.  The essence of the action is the above.} with the Zinn terms needed to make it satisfy the \PB
\be
{{\cal A}_{\rm Complete}}
=
{\cal A}_{\rm Yang\; Mills} 
=\int d^4 x 
\lt \{
-\fr{1}{4}
G^{a \m \n} 
G^{a}_{ \m \n} 
+
\S^{a \m} D_{\m}^{ab}
\w^{b}
- \fr{1}{2} K^a f^{abc}\w^b \w^c
\rt \}
\la{YMaction1}
\ee
where
\be
G^{a}_{\m \n} 
=\pa_{\m} A^a_{\n} 
-\pa_{\n} A^a_{\m} 
+ f^{abc} A^b_{\m}  A^c_{\n} ;
 D^{ab}_{\m} \w^b=
 \pa_{\m} \w^a
+ f^{abc} A^b_{\m}  \w^c
\ee

  The result is not unique however.  Any Yang-Mills theory, plus any number of U(1) gauge fields, will do as the solution.

If we had added an \OI\ here, the theory would not have changed in a fundamental way.  The \PB\ would still yield zero if one takes an action like
\be
 {\cal A}_{1}= {\cal A}_{\rm Free}
+ {\cal A}_{\rm Ordinary\;Invariant}
\ee
because
\be
{\cal P}\lt [ 
{\cal A}_{1}
\rt ]
=
{\cal P}\lt [ 
{\cal A}_{\rm Ordinary\;Invariant}
\rt ]=0
\ee
holds when ${\cal A}_{\rm Ordinary\;Invariant}$ is free of Zinns, and that is the assumption which defines ${\cal A}_{\rm Ordinary\;Invariant}$.  So the steps that follow from (\ref{pofenotzero}) do not happen for the case of adding a ${\cal A}_{\rm Ordinary\;Invariant}$. 
Note that the theory (\ref{YMaction1}) has the same Poisson Bracket as the free theory, but the two are not related by a local canonical transformation, as is particularly clear from (\ref{fullmonte}).   More detail about the canonical transformation, and what happens when one adds terms like 
 ${\cal A}_{\rm Ordinary\;Invariant}$ can be found in
\ci{dixnucphys}.

So why is it useful to find yet another way to derive 
the \YM? This exercise demonstrates a number of 
things:
\ben
\item
The \EI\  (\ref{lorentzinvei}) that gives rise to \YM\ exists in the free gauge theory.
\item
The  \PB\ (\ref{poissonofextra}) that results from the action 
(\ref{eadded}) is not zero, and it is also not even a cocycle of $\d_{\rm Free}$.  To get it to be a cocycle  of $\d_{\rm Free}$, we need to set the expressions 
(\ref{jacobihere}) and (\ref{jacobihereagain}) to zero.  These are both equivalent to  the Jacobi identity. 
\item
Then we find that the \PB\ reduces to a boundary of 
 $\d_{\rm Free}$, whose \PB\ is zero, so that we recover the \YM.
\item
At any stage this could have developed a problem.  But it does not because \YM\ exists.  What happens when we do not know where we are going? 
\een

It would be interesting to see what happens with other theories.  Here is a list of possible candidates and some present progress:
\ben
\item
We will see in \ci{DASSM} how these ideas apply to chiral scalar rigid SUSY. Many of the above steps will occur again, but the theory that emerges is new and there are an infinite number of \EI s in that theory,
and the constraints lead us to something rather like the Supersymmetric Standard Model.   So this is an example where the above procedure may lead us to something that is both useful and currently unknown.
\item
The Super Yang Mills theory in D=3+1 also has plenty of \EI s, and work needs to be done to analyze them.
\item
It seems likely that the theory of gravity would work much like the \YM, except that there must be interesting differences. 
\item
What happens here with the various theories of supergravity?   Could they possibly admit new couplings 
through this mechanism? 
\item
What happens with free theories of higher spin fields with multiple indices and more than two derivatives in their kinetic terms?  These seem to be needed to go beyond the analysis in  \ci{DASSM}.

\een

This exercise has not generated a new theory in this paper, but the method introduces a new point of view.  It is a different way to arrive at the Yang-Mills theory.  The \ELI\ 
 (\ref{lorentzinvei}) 
is a `seed' from which we can `grow' \YM\ by adding the `seed' to the action and then completing the action so that it satisfies the \PB.  The fact that the procedure works so nicely to generate  \YM\ seems to indicate that one should see what happens in other theories, using the same notions. There are many possible candidates for that. Any free  quantum field theory with a symmetry will do, and no sophisticated methods are needed to examine the low dimensional local BRST cohomology to look for \EI s.

\vspace{.2in}

\begin{center}
 {\bf Acknowledgments}
\end{center}
\vspace{.2in}
  I thank  Carlo Becchi, Philip Candelas, Rhys 
Davies,  Paul Frampton, John Moffat,  Peter Scharbach, Raymond Stora,  Xerxes Tata and
J.C. Taylor  for stimulating correspondence and conversations, and the Perimeter Institute for hospitality.

{\tiny \articlenumber}
%\tableofcontents

\end{document}

%% file: Johnmacros15.tex
\newcommand{\YM}{Yang-Mills theory}

\newcommand{\ELI}{Extraordinary Lorentz Invariant}
\newcommand{\dF}{\d_{\rm Free}}

\newcommand{\OI}{Ordinary Invariant}
\newcommand{\LIP}{Local Integrated Polynomial}
\newcommand{\EI}{Extraordinary Invariant}

\newcommand{\PB}{BRST Poisson Bracket}

\newcommand{\dB}{\d_{{}_{{}_{\rm BRST}}}}

\newcommand{\SM}{Standard Model}

\newcommand{\LT}{\LaTeX}

\newcommand{\bt}{\begin{tabular}{c}}
\newcommand{\et}{\end{tabular}}

\newcommand{\eb}{\ee\be } 
\newcommand{\ebp}{\rt.\ee\be\lt.} 
\newcommand{\bmat}{\lt ( \begin{array} }
\newcommand{\emat}{  \end{array} \rt )}

\newcommand{\ED}{\end{document}}

\renewcommand{\d}{\delta}

\newcommand{\m}{\mu}
\newcommand{\n}{\nu}	
\newcommand{\x}{\xi}

\newcommand{\s}{\sigma}

\newcommand{\w}{\omega}

\renewcommand{\S}{\Sigma}

\newcommand{\la}{\label}
\newcommand{\ci}{\cite}

\newcommand{\ds}{\documentstyle}	
\newcommand{\fr}{\frac}

\newcommand{\pa}{\partial}

\newcommand{\be}{\begin{equation}}
\newcommand{\ee}{\end{equation}}
\newcommand{\ba}{\begin{array}} 
\newcommand{\ea}{\end{array}}
\newcommand{\bea}{\begin{eqnarray}}
\newcommand{\eea}{\end{eqnarray}}

\newcommand{\Ra}{\Rightarrow}

\newcommand{\lt}{\left}
\newcommand{\rt}{\right}

\newcommand{\ben}{\begin{enumerate}}
\newcommand{\een}{\end{enumerate}}
\newcommand{\bitem}{\begin{itemize}}
\newcommand{\eitem}{\end{itemize}}

%% file: 3675.YMExtraInv.bbl
\begin{thebibliography}{99}
\bibitem{Bootstrapref} This amusing  old idea for the origin of the S-matrix is also closely related to the origin of supersymmetry through strings and duality. The notion of a Bootstrap in particle physics was popularized by G. Chew and others in the 60's and 70's.
\end{thebibliography}
